# A single quantum cannot be teleported

Daniele Tommasini

*Departamento de Física Aplicada, Universidad de Vigo, 32004 Ourense, Spain*

**Due to the Heisemberg uncertainty principle, it is impossible to design a procedure which permits perfect cloning of an arbitrary, unknown "qubit" (the spin or polarization state of a single quantum system)[1,2]. However, it is believed that a perfect copying protocol can be achieved, at least in principle, if the qubit to be copied is destroyed in the original system. Quantum teleportation[3,4] is supposed to allow for such a result. Here, this belief is shown to be invalidated by a fundamental uncertainty about the number of particles involved in any process, as predicted by Quantum Field Theory. As a result, teleportation cannot provide an infallible copying procedure for the single qubits, not even in the limit of perfect experimental sensitivities. The no-cloning theorem[1,2] can then be generalized to the case of destroying the original. Teleportation remains an interesting statistical procedure, having an unavoidable theoretical error at the percent level or few orders of magnitude smaller, depending on the physical process that is used. Although it cannot be made arbitrarily small, such an error is small enough to remain hidden in present experiments.**

Dated: 10/9/02

The problem which is addressed by teleportation is defined as follows[3]: one observer, called "Alice", is given a quantum system such as a photon or spin-$\frac{1}{2}$ particle, prepared in a state $|f\rangle$ unknown to her, and she wishes to communicate to another observer, "Bob", sufficient information about the quantum system for him to make an accurate copy (i.e. an exact replica). Of course, if Alice knew the state $|f\rangle$, she could just send a classical message with this information. However, in general $|f\rangle$ could only be determined experimentally if an infinite ensemble of identically prepared copies of the system were available, while we are supposing that Alice only has a single system. In general, if the relevant state space is not trivially one-dimensional, no measurement will give sufficient information to prepare a perfectly accurate copy of any arbitrary $|f\rangle$. In order to overcome this limitation, Alice and Bob share a pair of "ancillary" particles in an Einstein-Podolsky-Rosen[5] (EPR) entangled state. When Alice performs a complete set of compatible measurements on the original system and her ancillary particle, the other ancillary particle at Bob's is supposed to be put in an exact copy of the state $|f\rangle$ up to a unitary transformation, i.e. a matrix, which is determined from the result of Alice's measurement and can be communicated by a classical message to Bob[3]. Since the state of the original system has been destroyed by Alice's measurements, the possibility of such an infallible replica of the unknown state does not contradict the no-cloning theorem[1,2].

To be concrete, let me first consider the case which was considered explicitly by Bennet et al.[3] Let then the original system be a spin-$\frac{1}{2}$ particle, which will be called particle 1, and $|f\rangle = a|\uparrow_1\rangle + b|\downarrow_1\rangle$ be its spin state, in terms of the usual basis of the "up" and "down" eigenstates of, say, the z component of the spin. Let the EPR ancillary particles 2 and 3 also be spin-$\frac{1}{2}$ particles. Teleportation is based on assuming that the latter particles are created in a EPR-Bohm[5,6] spin-singlet state:

$$\left|\Psi_{23}^{(-)}\right\rangle \equiv \frac{1}{\sqrt{2}}\left(\left|\uparrow_2\right\rangle\left|\downarrow_3\right\rangle - \left|\downarrow_2\right\rangle\left|\uparrow_3\right\rangle\right). \tag{1}$$

Now, Alice chooses to perform a complete set of compatible measurements that project the state of particles 1 and 2 on the "Bell basis" $\left|\Psi_{12}^{(\pm)}\right\rangle \equiv \frac{1}{\sqrt{2}}\left(\left|\uparrow_1\right\rangle\left|\downarrow_2\right\rangle \pm \left|\downarrow_1\right\rangle\left|\uparrow_2\right\rangle\right)$, $\left|\Phi_{23}^{(\pm)}\right\rangle \equiv \frac{1}{\sqrt{2}}\left(\left|\uparrow_1\right\rangle\left|\uparrow_2\right\rangle \pm \left|\downarrow_1\right\rangle\left|\downarrow_2\right\rangle\right)$. On the other hand, the state of the three particle system is assumed to be the tensor product

$$\left|\Psi_{123}\right\rangle = \left|f\right\rangle\left|\Psi_{23}^{(-)}\right\rangle = \frac{1}{2}\left[\left|\Psi_{12}^{(-)}\right\rangle\left(-a\left|\uparrow_3\right\rangle - b\left|\downarrow_3\right\rangle\right) + \left|\Psi_{12}^{(+)}\right\rangle\left(-a\left|\uparrow_3\right\rangle + b\left|\downarrow_3\right\rangle\right) + \left|\Phi_{12}^{(-)}\right\rangle\left(a\left|\downarrow_3\right\rangle + b\left|\uparrow_3\right\rangle\right) + \left|\Phi_{12}^{(+)}\right\rangle\left(a\left|\uparrow_3\right\rangle - b\left|\downarrow_3\right\rangle\right)\right] \tag{2}$$

With these assumptions, if Alice gets, say, $\left|\Psi_{12}^{(-)}\right\rangle$ as the result of her measurements on particles 1 and 2, and she send a classical message to Bob with this information, Bob will know with certainty that (up to an irrelevant minus sign) the state of particle 3 will be precisely $\left|f\right\rangle = a\left|\uparrow_3\right\rangle + b\left|\downarrow_3\right\rangle \equiv \begin{pmatrix} a \\ b \end{pmatrix}$; if Alice gets $\left|\Psi_{12}^{(+)}\right\rangle$, then Bob will get the state $\left(-a\left|\uparrow_3\right\rangle + b\left|\downarrow_3\right\rangle\right) = \begin{pmatrix} -1 & 0 \\ 0 & 1 \end{pmatrix}\left|f\right\rangle$, etc. In other words, using the classical message carrying the information on the result of Alice's measurements (telling which of the four states $\left|\Psi_{12}^{(\pm)}\right\rangle$, $\left|\Phi_{12}^{(\pm)}\right\rangle$ she has observed, which amounts to two bits of information), Bob will have his particle 3 put in the state $\left|f\right\rangle$, up to a unitary transformation. The state $\left|f\right\rangle$ is then said to be "teleported" from particle 1 to particle 3 (from Alice to Bob). The previous scheme can be easily generalized to EPR pairs made of photons, kaons or other particles that have two spin or polarization states.

The above is the usual presentation of teleportation, which is supposed to provide Bob with "an exact replica of the unknown state $\left|f\right\rangle$ which Alice destroyed"[3]. In other words, it is believed that "teleportation of a polarization state can occur with certainty in principle"[7] (equivalent sentences can be found in almost all the papers dedicated to the subject). As we have seen, this conclusion is based on the hypothesis that Alice and Bob can actually share a pair of EPR particles described by the entangled state of eq. (1), at least in principle. However, this assumption is invalidated by the modern Quantum Field Theory (QFT) description of Particle Physics. Although the two particle EPR-Bohm state belongs to the state space, it cannot be produced by any physical process.

As we shall see, a generalization of the uncertainty principle for the relativistic regime, as described by QFT, makes it impossible to produce a state with a definite number of particles. This corresponds to a basic characteristic of the QFT description of Particle Physics[8]: *it predicts a non-vanishing and finite probability for any process that does not violate the fundamental symmetries.* In fact, in Particle Physics experiments, it is so rare to find an "accidental cancellation" for the rate of an allowed process that such a case would be considered a hint for some new symmetry forbidding that channel. In other words, all the new particles that can be created without violating the universal conservation laws can actually be produced, and any definite process involving the creation of a particular set of particles has its corresponding amplitude of probability, which can eventually be computed approximately by drawing the relevant Feynman diagrams[8] (when perturbation theory is applicable). In particular, since photons have zero rest mass, their energy can be arbitrarily low. Since they also have zero charge and colour, we can conclude that *an arbitrary number of additional photons, with total energy compatible with energy conservation, can always be created in coincidence with any physical process.*

Let us now evaluate the amount of such an uncertainty. First, in the EPR-Bohm case, in which the considered EPR pair are spin-$\frac{1}{2}$ particles, additional photons can *always* be radiated *at least* by the external legs of the Feynman diagram[8] that describes the process, as shown in Fig. 1. The rate for the process involving an additional photon is suppressed merely by a factor of order $a$, where $a \cong 1/137$ is the fine-structure constant, with respect to the rate for producing only the EPR pair. In fact, a similar correction also applies when the EPR-Bohm particles have no total electric charge. For instance, neutral kaons, previously considered as possible EPR particles, are made out of charged quarks[9]; since the production process involves such constituent quarks, the order of magnitude for the rate of the process involving additional photons is suppressed merely by a factor $a$, as compared to the rate of the process without any additional photon, as in the case of charged fermions. In the case of an EPR pair of photons, two additional photons can also be radiated by the external legs of the relevant production Feynman diagram through loop diagrams involving charged fermions. At low energies, the rate for such a loop diagram is small, although it is finite and cannot be made *arbitrarily* small; more importantly, even in this case there is a much larger rate for the emission of additional photons due to the charged particles that are necessarily involved in the relevant production vertex. Such a rate is usually of the order $a$ or (in the presence of selection rules) of the order $a^2$, as compared to the rate for the production of the EPR pair alone.

To summarize, the fundamental uncertainty about the number of particles makes it impossible to produce a pair of particles in the EPR state of Eq. (1) as a result of a physical process. Such a state can be a useful approximation within an error which may be reduced to a fraction of the percent level, but which cannot be made arbitrarily small.

To illustrate this, I will now explicitly consider few particular processes to produce an entangled EPR pair, and estimate the amount of the indetermination on the number of particles, and the effect on teleportation.

The first example is the decay of the neutral pion $p^0$. This is a good elementary way to obtain a pair of entangled photons having vanishing total angular momentum, since $p^0 \rightarrow gg$ is by far the most probable decay channel of the pion. Moreover, this example has the advantage that we can know the numerical amount of the uncertainty in the number of particles by using the existing data for the branching ratios of the decay channels of the pion. Heuristically, we can say that the state that is produced as a result of the decay of an ensemble of $p^0$s is

$$\left|\Psi\right\rangle \approx 0.9940\left|y_{gg}\right\rangle + 0.1095\left|y_{ge^+e^-}\right\rangle + 0.0056\left|y_{e^+e^-e^+e^-}\right\rangle + 0.00027\left|y_{e^+e^-}\right\rangle + ... \quad (3)$$

where each component $\left|y_{gg}\right\rangle$, $\left|y_{ge^+e^-}\right\rangle$, $\left|y_{e^+e^-e^+e^-}\right\rangle$, $\left|y_{e^+e^-}\right\rangle$,... respectively describes a normalized state involving two photons, or a photon and an electron-positron pair, etc., and the coefficients are the square root of the rates for the corresponding channels, as given in the Review of Particle Physics[9]. The previous simple description can be used for the present purposes, leading to the same results that can be obtained by a more correct treatment in terms of density matrices or QFT Green functions[8]. In this particular case, the branching ratio for the production of (two) additional photons is exceptionally small, being of the order of $a^4$. The main effect of the uncertainty in the number of particles is then due to the component $\left|y_{ge^+e^-}\right\rangle$.

In fact, suppose that the products of the pion decay are used as the EPR particles 2 and 3 in a teleportation procedure. When Alice receives a photon (particle 2), she performs her Bell measurements on particles 1 and 2, and communicates her results with a classical message to Bob. There is then a probability of $0.994^2 = 98.8\%$ that Bob will receive another photon (that would be particle 3) as described by the entangled state component $\left|y_{gg}\right\rangle$. In such a fortunate case, the total angular momentum of the two EPR photons will sum to zero, the value of the spin of the decaying pion, and the teleportation protocol would seem to work; however, in the remaining 1.2% of the cases, *Bob will not receive a photon entangled with Alice's,* and the single qubit of the original system (particle 1) remains destroyed by Alice's measurement without being teleported. It is very important to note that this unavoidable loss is due to a theoretical uncertainty, and not to experimental limitations. This is analogous to the no-cloning theorem[1,2], that forbids perfect replication on a purely theoretical basis. By chance, it might be that the single qubit that is under consideration is teleported, but in a general single event we cannot know that this will be the case, since we would be randomly wrong in 1.2% of the cases. Therefore, teleportation as a copying procedure cannot be perfect, contrary to current belief. Actually, due to this fundamental uncertainty, the usual assumption that a state (qubit) can be attributed to the single system rather than to the ensemble also becomes questionable[10].

A class of mechanisms that has been proposed for the production of EPR pairs is based on the excitation and subsequent decay of a bound system such as an atom. For instance, an atom is excited to a state $b$ of energy $E_b$ which then decays to the ground state in two steps passing through an intermediate state $c$ of energy $E_c$, thus emitting two photons in cascade, the first with energy $E_b - E_c$ and the second with energy $E_c - E_g$, where $E_g$ is the energy of the ground state $g$ (the energies of both photons being defined within a width depending on the lifetimes of the excited levels). According to this picture, if the excited state $b$ has the same angular momentum as the ground state $g$, the two photons are emitted in an entangled state of zero total angular momentum. However, at higher orders in perturbation theory, any of the two decaying steps can proceed through multiphoton emission[11]. Since the transitions from state $b$ to $c$ and from $c$ to $g$ are due to electric dipole matrix elements[11], the next leading order in perturbation theory involves the emission of two additional (electric-dipole) photons. For instance, instead of emitting a photon with energy $E_b - E_c$, the first step of the decay of the atom can proceed through the emission of three photons having the same total energy. The rate for such a process is suppressed merely by a factor $a^2$ as compared to the rate for the production of the EPR pair alone. Heuristically, we can say that the state which is produced by the decay of the excited atom is

$$\left| \Psi \right\rangle \approx c_1 \left| y_{gg} \right\rangle + c_2 \left| y_{gggg} \right\rangle + ... \tag{4}$$

where $\left| c_2 \right|^2 \approx o(a^2) \approx o(10^{-6} - 10^{-4})$, and $\left| c_1 \right|^2 \cong 1 - \left| c_2 \right|^2$. Therefore, Alice and Bob will share an EPR pair of correlated photons with a probability $\left| c_1 \right|^2$. However, there is a probability $\approx \left| c_2 \right|^2$ that Alice will get a photon from the EPR source and perform her Bell measurements on it and on her qubit, while Bob will either not receive anything (conservation of momentum does not force a second photon to reach Bob, in the presence of additional photons), or (less probably) will receive a photon that is not anticorrelated to that used by Alice (in this case, the angular momentum is conserved for the four photons produced by the EPR source, not for a couple of them). This implies an error in teleportation of the order of $10^{-6} - 10^{-4}$. Similar considerations can be repeated for the technique of parametric down conversion[12,13] that is used in actual teleportation experiments[14,15,16,17].

To conclude, I have proved that teleportation does not allow for a certain copying procedure, always having a finite non-vanishing theoretical error, typically in the range $10^{-6} - 10^{-2}$. There is no guarantee that a single qubit will be successfully teleported. This generalizes the no-cloning theorem[1,2] to this case of destroying the original. On the other hand, the sub-ensemble of the coincident events (corresponding to Alice and Bob both receiving their ancillary particle) has a very small probability for the production of additional particles. If those events are selected after two local measurements by both Alice and Bob, the resulting *a posteriori teleportation*[18] may be almost perfect, at least in principle. Note however that this concept, requiring a feedback communication from Bob, is conceptually different from teleportation, which is supposed to provide (in principle) a certain copy of any arbitrarily given qubit. In other words, if the process is repeated on a great number of qubits, a fraction of them of the order $10^{-6} - 10^{-2}$ will be lost, while for the remaining events *a posteriori teleportation* may be considered to be successful with very good accuracy. On the other hand, the fact that two local measurements on both particles of an EPR pairs are needed to ensure almost perfect correlations, can have important implications in the interpretation of the quantum theory, removing one of the supposed proofs of the existence of an instantaneous influence between distant measurements[10]. In fact, the supposed possibility of teleporting a single qubit has been considered a proof of the apparent "nonlocality" of the quantum theory[17]. We see now that this proof is also removed.

1. Wootters, W. K., and Zurek, W. H., A single quantum cannot be cloned, *Nature* **299,** 802-803 (1982).

2. Ghirardi, G. C., and Weber, T., *Nuovo Cimento* **78B,** 9 (1983).

3. Bennet, C. H., *et al.,* Teleporting an unknown quantum state via dual classical and Einstein-Podolsky-Rosen channels, *Phys. Rev. Lett.* **70,** 1895-1899 (1993).

4. Sudbery, T., Instant teleportation, *Nature* **362,** 586-587 (1993).

5. Einstein, A., Podolsky, B., and Rosen, N., Can quantum mechanical description of physical reality be considered complete?, *Phys. Rev*. **47,** 777-780 (1935).

6. Bohm, D., *Quantum Theory,* Prentice-Hall (1951).

7. Y. H. Kim, S. P. Kulik and Y. Shih, Quantum teleportation of a polarization state with a complete Bell state measurement, *Phys. Rev. Lett.* **86,** 1370-1373 (2001).

8. Weinberg, S., *The quantum theory of fields,* vols. I and II, Cambridge University Press (1996).

9. Hagiwara, K., et al., The review of Particle Physics, *Phys. Rev. D* **66,** 010001 (2002).

10. Tommasini, D., Reality, measurement and locality in Quantum Field Theory, *JHEP* **0207,** 039 (2002).

11. Cohen-Tannoudji, C., Dupont-Roc, J., and Grynberg, G*., Atom-photon interactions,* John Wiley & Sons (1992).

12. Kwiat, P. G., *et al.,* New high-intensity source of polarization-entangled photon pairs, *Phys. Rev. Lett.* **75,** 4337-4341 (1995).

13. Barnes, W., Survival of the entanglement, *Nature* **418,** 281-282 (2002).

14. Bouwmeester, D., *et al.*, Experimental quantum teleportation, *Nature* **390,** 575-579 (1997).

15. Boschi, D., *et al.*, Experimental realization of teleporting an unknown state via dual classical and Einstein-Podolsky-Rosen channels, *Phys. Rev. Lett.* **80,** 1121-1125 (1998).

16. Furusawa, A., *et al.*, E. S., Unconditional quantum teleportation, *Science* **282,** 706-709 (1998).

17. Sudbery, T., The fastest way from A to B, *Nature* **390,** 551-552 (1997).

18. Braunstein. S. L., and Kimble, H. J., *A posteriori* teleportation, *Nature* **394,** 840-841 (1998).

Acknowledgements

I thank H. Michinel, G. C. Ghirardi, A. Aste, R. Schnabel and G. L. Celardo for useful discussions, and A. Tommasini, C. Fernández, E. Carballo and R. Ramanathan for help.

**Correspondence should be addressed to the author (e-mail: daniele@uvigo.es)**

**Figure caption**

**Figure 1** Feynman diagram describing the production of an EPR pair of charged spin-½ particles, A and B, in coincidence with four additional photons (wavy lines). The dashed "blob", representing the particular process and initial particles that are considered, can also radiate additional photons (one in the figure).

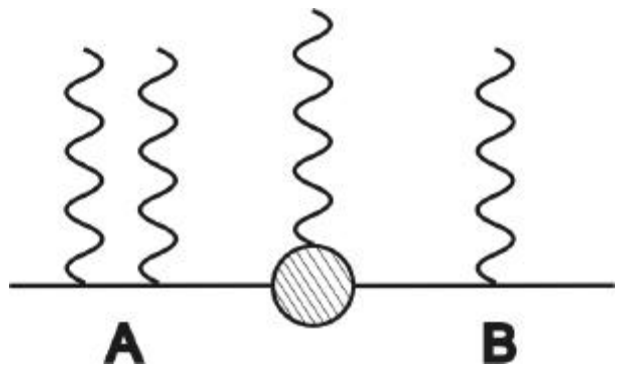

**tommasini_fig1**